# Nanometer Resolution Elemental Mapping in Graphene-based TEM Liquid Cells

*Daniel J. Kelly,[†,§] Mingwei Zhou[‡,§], Nick Clark[†,§], Matthew J. Hamer[‡,§], Edward A. Lewis[†],*

*Alexander M. Rakowski[†,§], Sarah J. Haigh\*[†,§], and Roman V. Gorbachev\*[‡,§]*

[†]School of Materials, University of Manchester, Oxford Road, Manchester, M13 9PL, UK

[‡]School of Physics and Astronomy, University of Manchester, Oxford Road, Manchester, M13 9PL, UK

[§]National Graphene Institute, University of Manchester, Oxford Road, Manchester, M13 9PL, UK

We demonstrate a new design of graphene liquid cell consisting of a thin lithographically patterned hexagonal boron nitride crystal encapsulated from both sides with graphene windows. The ultra-thin window liquid cells produced have precisely controlled volumes and thicknesses, and are robust to repeated vacuum cycling. This technology enables exciting new opportunities for liquid cell studies, providing a reliable platform for high resolution transmission electron microscope imaging and spectral mapping. The presence of water was confirmed using electron energy loss spectroscopy (EELS) via the detection of the oxygen K-edge and measuring the thickness of full and empty cells. We demonstrate the imaging capabilities of these liquid cells by tracking the dynamic motion and interactions of small metal nanoparticles with diameters of 0.5-5 nm. We further present an order of magnitude improvement in the analytical capabilities compared to previous liquid cell data, with 1 nm spatial resolution elemental mapping achievable for liquid encapsulated bimetallic nanoparticles using energy dispersive X-ray spectroscopy (EDXS).

One of the most attractive and unique capabilities of the scanning transmission electron microscope (STEM) is its ability to perform high spatial resolution elemental analysis through EDX and EEL spectroscopies. Nevertheless, a key limitation for this technique is the requirement for high vacuum conditions to achieve the best imaging and analysis. Several studies have revealed that the structure of functional materials at room temperature in a vacuum may be significantly different from that in their operational environment.[1,2] In situ electron microscopy has emerged as a solution to allow imaging to be performed under more realistic environmental conditions. Unfortunately creating an in situ environment within the TEM has only been achievable by sacrificing some of the instruments spatial resolution imaging and elemental analysis capabilities.[3,4]

To study a liquid sample in the electron microscope without drying or freezing, the specimen is contained inside a liquid cell; an approach which has provided valuable insight into many dynamic processes across biology, chemistry, geology, and materials science.[5–8] The liquid cell is created by capturing a thin layer of solution between two impermeable, but electron transparent membranes which prevent evaporation into the vacuum of the microscope. The electron beam passes through both the membranes and the specimen to form the image. Commercially available liquid cells are composed of microfabricated silicon nitride membrane windows (20-50 nm thick) that are electron transparent yet capable of withstanding the considerable pressure difference.[8,9] This type of cell has been used to great effect; for instance entire biological cells have been studied in their native environment without the need for drying or freezing, which could produce artefacts,[10,11] live electrochemical reactions can be monitored at high resolution in real time,[12] and the beam-liquid interaction can been exploited to study the growth of metallic nanocrystals from solution.[13–15]

Liquid cell STEM is the only technique with the potential to directly probe elemental distributions in liquids at atomic resolution. However, current designs of liquid cells have several limitations which need to be overcome to make this possible. The primary issue is excessive beam scattering in both the silicon nitride membranes and the liquid media, which limits the spatial resolution achievable for both imaging and analysis. The exact depth of liquid layer in the cell is difficult to control accurately and often varies across different parts of the cell due to membrane bowing. There is evidence that this can cause changes to the behavior of the system, for example Brownian motion is suppressed in very thin liquid samples.[16] In addition, the geometry of many liquid cell designs can prevent X-rays emitted from the specimen from reaching the detectors, reducing the quality of EDX spectrum imaging.[17]

In order to improve the imaging resolution achievable with conventional liquid cells, SiN windows have been replaced with graphene; chosen due to its outstanding mechanical properties,[18,19] physical impermeability[19,20] and chemical stability. Suspended graphene membranes up to ten micrometers in diameter can be routinely fabricated and have the further advantage of greatly reducing deleterious beam induced charging effects[21–23] since graphene is an excellent conductor of electricity[24] and heat.[25] Graphene is also chemically inert in the absence of defects or can be functionalized to make it hydrophobic.[26] Hermetic sealing of the cell is made possible by a strong van der Waals interaction between graphene and other atomically flat surfaces.[27]

To date, there have been several successful demonstrations of TEM imaging in graphene liquid cells (GLC), primarily based on the prototype design created by Yuk et al.,[28] where cells are fabricated by bringing together two sheets of CVD graphene whilst submerged in a liquid.[28–31] On contact, van der Waals forces act to maximize the contact area between the two sheets, forcing small amounts of trapped liquid into micro- or nanometer-scale pockets. This is conceptually simple but has clear disadvantages: the formation of the pockets is random so the heterogeneity, volume, location, and thickness of the encapsulated liquid cannot be predetermined or controlled. The locations useful for TEM are therefore hard to find as well as unstable under electron beam illumination due to the fragility of the polycrystalline CVD graphene.[32,33] More importantly the hydrostatic pressure in such bubbles has been reported to reach up to 1 GPa, a drastic difference that is expected to significantly modify most chemical processes compared to ambient conditions. This pressure varies over two orders of magnitude depending on the pocket dimensions and its exact value is hard to determine from TEM images alone.[34] In addition, this design concept offers no obvious route to further technical advancement, such as the addition of electrochemical, heating, or flow and mixing capabilities which are invaluable fixtures in the field of in situ electron microscopy as it stands.

An advanced technique has been reported where cells are fabricated by etching cylindrical holes into a silicon nitride membrane and encapsulating it with monolayer graphene under liquid.[35] While offering control of the cell dimensions and density, individual liquid pockets were found to dry out after ~10 minutes[35] of TEM imaging often causing all adjacent cells to lose liquid as well. This leakage is likely due to the roughness of the SiN surface preventing a complete seal with graphene and causing liquid diffusion within the interface.

In this work we present a new engineered graphene liquid cell (EGLC) design based on a van der Waals heterostructure platform,[36] where top and bottom graphene windows are separated by a thin layer of hexagonal boron nitride (hBN). This approach offers unprecedented control of the cell dimensions and a completely leak-tight liquid enclosure that is stable under prolonged STEM imaging. We demonstrate that our engineered liquid cell design provides new opportunities for probing liquid phase reactions without the need to compromise capabilities for nanometer resolution elemental mapping. We show an order of magnitude improvement in the elemental mapping with the record of ~1 nm spatial resolution achieved on complex metallic nanostructures in water using STEM EDX spectrum imaging.

Sample fabrication starts with selection of a thin hBN crystal exfoliated on an oxidized silicon wafer. Depending on the required liquid cell depth the hBN thickness can be selected to be only a few atomic layers or up to several micrometers. We then create a regular array of circular holes in the crystal using a lithographically defined reactive ion etching process. After annealing the crystal to remove resist residue we pick it up with the top graphene crystal using the stacking technique described in Kretinin et al.[37] The resulting stack is then deposited onto the bottom graphene layer while submerged in a liquid media, creating perfectly sealed cylindrical "wells" as depicted in Figure 1a and with a HAADF STEM top-down view shown in Figure 1b.

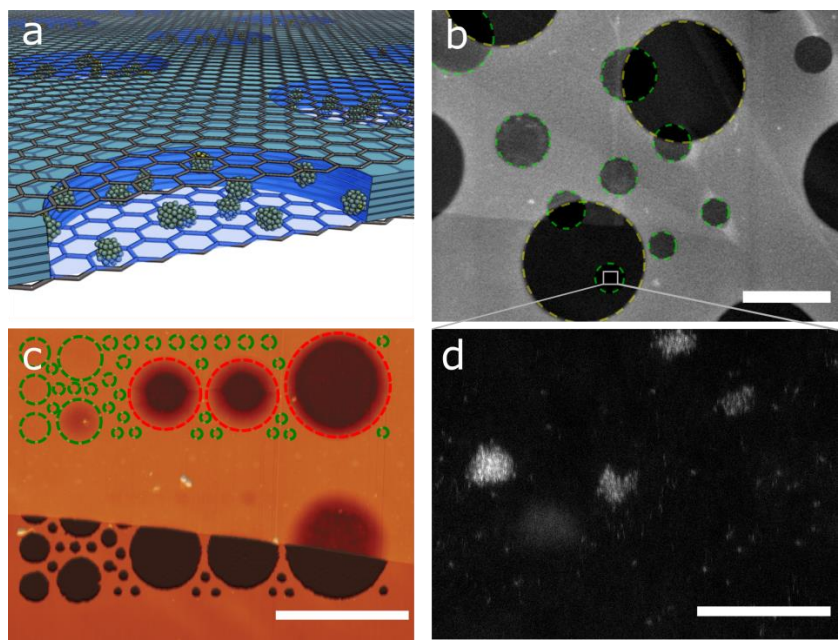

**Figure 1.** (a) An illustration of the engineered graphene liquid cell configuration containing nanocrystals (not to scale). (b) HAADF STEM image showing the EGLC wells (outlined in green) overlapping the holes of the quantifoil TEM support grid (outlined in orange). (c) AFM mapping of the EGLC supported on a silicon wafer, showing filled wells (outlined in green) and empty wells (outlined in red). The edge of the top graphene flake is clear, with the lower part of the image showing the underlying patterned hBN spacer crystal. (d) Typical HAADF STEM image of atomically resolved Pt nanoparticles, precipitated from 0.35 mM $H_2ClPt_6$ solution, inside a graphene well, imaged at 200 kV. Scale bars are (b) 1 µm (c) 2 µm (d) 5 nm.

Topographical AFM imaging can be used to assess the effective filling of the wells during fabrication. Figure 1c shows a group of wells containing liquid outlined in green, in contrast to empty wells shown outlined in red. For the empty wells the AFM measures a depth of 30 nm in the center of the well relative to the surrounding spacer, a height equal to the thickness of the hBN spacer crystal for this sample. In contrast, in the center of filled wells the graphene only sags by 1 - 5 nm depending on the cell diameter. The absence of significant curvature of the upper graphene sheet indicates that the hydrostatic pressure of the trapped liquid is close to 1 bar,[19,34,38] providing realistic conditions for many potential applications. The geometry of wells can be customized for a particular experiment but our typical designs have over 1000 densely packed circular wells ranging from 100 nm to 2 µm in diameter of which at least half have no defects and nearly identical topography. The edge of the top graphene window is seen in the lower part of Figure 1c and below this the full pattern of the underlying hBN spacer is clearly visible. Repetitive AFM measurements performed after 26 hours of vacuum exposure ($10^{-7}$ mbar) show that the liquid cells remain filled, confirming the absence of any leakage as a result of vacuum cycling.

We found that using 2 and 3 layer-thick graphene for windows makes fabrication yield and electron imaging stability significantly better than monolayer material, with no noticeable decay in STEM imaging or spectroscopy resolution. During an extensive performance evaluation we exposed some liquid cells to temperatures above 120 °C causing expansion of encapsulated liquid (IPA and water). The resulting pressure reaches 116 ± 46 bar, as estimated from membrane bowing[19,39] (see SI) with no apparent leakage of the liquid. Graphene's high elastic modulus[18] is potentially highly beneficial for TEM imaging of liquid media as it means a small change in volume leads to a large change in the cell pressure, which could suppress bubble formation.

To image the liquid inside the cells the EGLC graphene-hBN-graphene stack is transferred onto a TEM support with a regular array of holes (for detailed fabrication information see SI). A high angle annular dark field (HAADF) STEM image (Figure 1b) reveals an over view of the wells showing where several liquid wells (green) overlap with holes in the quantifoil TEM support grid (yellow). The HAADF signal scales with atomic number so the darkest areas are thinnest and most suitable for high resolution electron imaging, containing just two graphene windows and the encapsulated liquid specimen. In this example the thickness of the hBN spacer is 30 nm and the diameters of the patterned holes are in the range 100 - 1500 nm, resulting in liquid cells with cylindrical volumes in the range 1 - 200 zL. Higher magnification HAADF STEM images (e.g. Figure 1d) reveal the presence of Pt nanoparticles within the liquid well. HAADF STEM imaging provides several advantages over the more commonly used TEM including high contrast for dense nanoparticles relative to a lower atomic number liquid phase, a higher resolution with respect to liquid thickness when imaging particles in liquid,[8] and control of local electron dose to only the area being imaged.[40] The use of a graphene window with an ultra-low scattering cross section and small liquid thickness (~30 nm) allows very high resolution imaging of the nanoparticles in solution, clearly resolving the atoms in nanoparticles less than 0.5 nm in diameter.

To demonstrate the excellent imaging capabilities achievable in our EGLCs compared to conventional SiN liquid cells, we studied the formation and growth dynamics of small tungsten nanoparticles precipitated from a saturated aqueous $WCl_6$ solution. Beam induced reduction of aqueous salts is a widely studied method used to gain insight into the nucleation and growth of metal nanoparticles in solution.[15,41] The electron beam instigates radiolysis of the encapsulated water, resulting in its decomposition and the propagation of a variety of radicals and reactive molecular species throughout the cell.[42,43] Among these species are aqueous electrons which can reduce soluble metal ions to form solid metal clusters, and ultimately nanocrystals, a process which can be monitored by TEM imaging at high resolution in real time.[14,42,43]

The majority of observed particles nucleate immediately during first few seconds of imaging[44] and undergo random diffusive motion more pronounced in smaller particles, similar to that reported by Zheng et al.[45] An example of particle tracking is shown in Figure 2a where trajectories of a few representative particles are overlaid on top of the first frame. The tracking data was obtained from a HAADF STEM image series (80 kV accelerating voltage, 110 pA probe current, 8 µs pixel dwell time, image size of 512 × 512 pixels) with 199 frames at 2.5 s/frame and corrected for specimen drift (details and video available in SI).

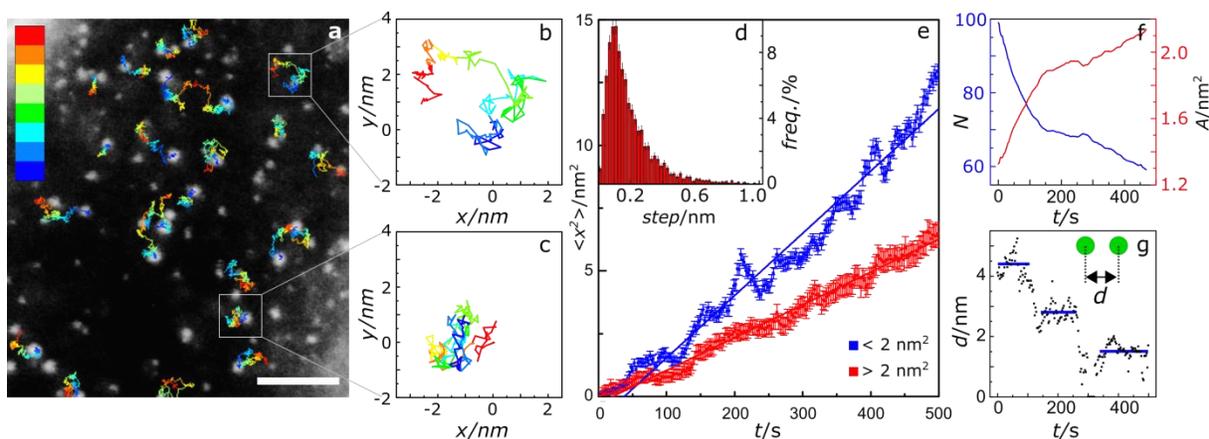

**Figure 2.** Tracking of nanoparticle motion for small tungsten nanocrystals in water (data extracted from the series of HAADF STEM images in Supplementary Video 1). (a) The first frame from the video with the motion of some of the individual particles superimposed. Scale bar is 10 nm. Trajectories of individual nanocrystals are overlaid with time (in seconds) represented by a color chart where blue is t = 0 s and each color block is a 60 second increment, (b, c) the movement paths for two individual nanocrystals with mean areas of (b) 1.3 and (c) 1.7 $nm^2$. (d) The relative frequency of different magnitudes of 'displacement per frame' (step) for all particles studied. (e) The mean square displacement as a function of time for different sizes of nanocrystal. The nanocrystals were separated in to two categories (>2 $nm^2$ and <2 $nm^2$). (f) Dual plot showing average nanocrystal projected area, A, and the population of nanocrystals, N,

detected per unit time, (g) the interparticle distance, d, plotted as a function of time for two individual nanocrystals exhibiting correlated motion prior to a coalescence event.

Figure 2e shows the mean square displacement, $\langle x^2 \rangle$ as a function of time, averaged within two groups of particles based on their size. A statistical analysis of ~5000 measured displacements for individual particles between neighboring frames is further provided as a histogram in Figure 2d, showing a single peak, centered at the most frequent value of 183 ± 4 pm (accuracy of the tracking and drift correction used was ~10 pm). The minimum observable particle displacement is of the order of 100 pm. In classical Brownian motion particle movement is driven by random momentum change due to collisions with atoms or molecules. Considering a tungsten nanoparticle with a projected area of 2 nm$^2$, one can estimate the collision rate with the surrounding water molecules as ~10$^{14}$ s$^{-1}$. Neglecting beam induced heating, which has been estimated to be only a few degrees for similar systems,[45] and considering average thermal energy at 300 K gives an estimated mean free path for each collision of <1 pm. The displacements we observe are on the order of 100 pm so are therefore clearly the result of many smaller steps, yet this simplistic calculation for a bulk system predicts much larger movement than we see experimentally. Nevertheless, the observed linear behavior is well described by 2D Brownian model where $\langle x^2 \rangle = 4Dt$, with smaller nanocrystals moving faster as expected. The resulting mean diffusion coefficients are calculated as D = 3.25×10$^{-3}$ nm$^2$ s$^{-1}$ for larger particles (with a measured projected area of greater than 2 nm$^2$, shown red in Figure 2e, mean size 2.84 nm$^2$ and standard deviation 0.44 nm$^2$) and D = 6.18×10$^{-3}$ nm$^2$ s$^{-1}$ for smaller particles (with a measured projected area of less than 2 nm$^2$, blue in Figure 2e, mean size 1.26 nm$^2$ and standard deviation 0.55 nm$^2$). These values are consistent with a previous observation of particles within graphene liquid cells,[28] but 10-100 times lower than those usually observed for SiN windowed liquid cells[45,46] and over $10^6$ times smaller than expected values for bulk water. The presence of surfaces is known to inhibit diffusion of particles in liquid and we hypothesize that this restricted motion can be explained by combined interactions of the nanocrystals with the graphene windows, water molecules, other nanocrystals, beam irradiation and hydrocarbon contamination in the cell.

The cells remain stable, retaining liquid even after several hours in vacuum. If a cell is punctured by focusing the 80kV electron probe at a single spot for a prolonged period, we are able to image in situ the motion of nanoparticles as water leaves the cell and evaporates into the TEM vacuum. The particles move rapidly as the liquid flows out, with a directional mean square displacement 4 times higher than the random motion of the nanocrystals in the intact cell (see Supplementary Video 2). Importantly, water is not lost from all other surrounding cells when one is burst, due to the van der Waals seal which isolates wells from their neighbors.

During the experiment we observe a significant decrease in the number of particles accompanied by an increase in their average size, caused by combined effects of Ostwald ripening (where larger particles grow at the expense of smaller particles) as well as particle coalescence (Figure 2f). Each coalescence event is preceded by a clear change in movement pattern of the two particles; they are seen to interact and exhibit correlated motion about each other over extended periods up to 100 s which ends abruptly with contact. In a typical example shown in Figure 2g, two particles 0.6 nm in diameter undergo diffusive motion while keeping their center-center distance between 2 and 3 nm for ~200 s seconds. The average separation distance varies for individual pairs, usually being integer multiples of a ~ 1 nm step. Due to the large amount of time taken for the particles to overcome this apparent barrier to coalescence, it is unlikely that the behavior is based on the particles aligning along specific crystal facets by simple rotation or reorientation. Instead it may be attributed to changes in particle structure or local environment, a behavior similar to that reported by Yuk et al.[28]

When two or more particles appear to come into close proximity in our projected images they either interact via coalescence or pass each other unaffected. The latter behavior can be explained by the particles moving at different depths in the liquid. These events are seen at similar rates, which can be explained if particles are localized on one of the graphene windows and are effectively exhibiting surface diffusion along the top or bottom graphene liquid interface. Indeed, for particles randomly distributed throughout 40 nm cell depth their collision within ~ 3 nm coalescence range would be an order of magnitude less likely. This hypothesis is supported by the good agreement between the particle motion measured in Figure 1e and the model for 2D Brownian motion.A further advantage of our EGLC design is its compatibility with STEM EDX and/or EEL spectrum imaging to gain elemental information at high spatial resolution. In most traditional silicon nitride liquid cells elemental analysis by EDX spectroscopy is challenging due to the penumbra of the holder which blocks characteristic X-rays from reaching the detector.[17] The spatial resolution of EEL spectroscopic analysis is also limited in such systems due to scattering induced by the thickness of the cell windows and large liquid cell volume, although Wang et al have shown the advantage of EEL spectroscopy in graphene liquid cells, high spatial resolution elemental mapping of specimens in liquid environment has yet to be achieved.[47] We have previously shown that modification of the SiN liquid cell design to minimize shadowing from the holder allows EDX elemental mapping to be performed with a spatial resolution of ~10 nm.[48]

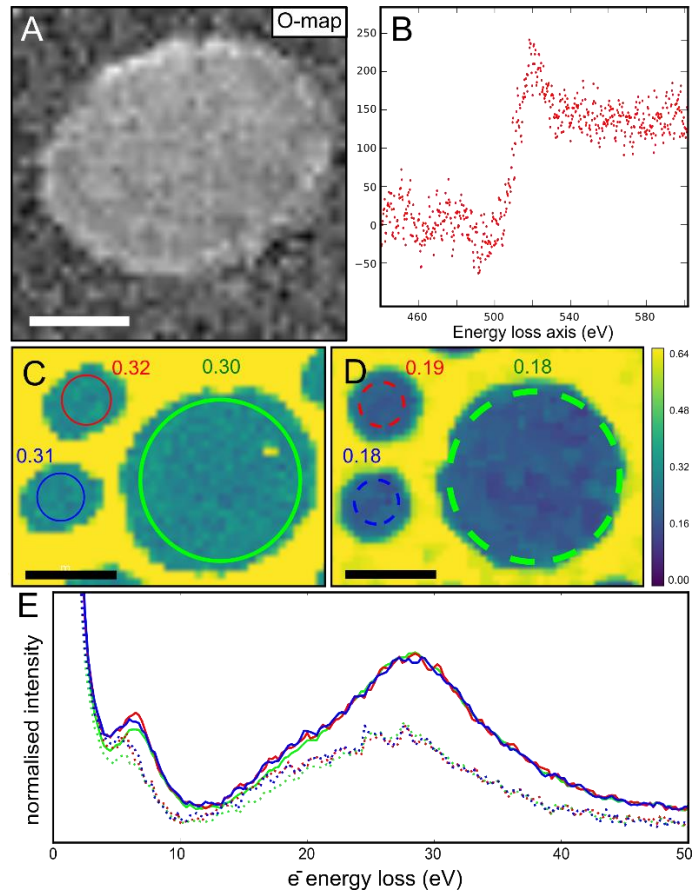

**Figure 3** EELS characterisation of an EGLC. (a) Mapping the oxygen K-edge for a EGLC with the extracted oxygen signal integrated over the cell shown in (b). (c),(d) Mapping the relative thickness (t/λ) of (c) filled and (d) empty cells respectively (obtained using the log-ratio method on the low-loss spectra[49]). The averaged thickness value for each hole is indicated. (e) The normalized low loss spectra integrated over the holes indicated in (c) and (d). Solid lines are full cells and dashed lines are empty cells. Scale bars are 100 nm.

In order to evaluate the full potential of our new cell design we have conducted elemental mapping using EDX and EEL spectroscopies. The presence of the oxygen K edge in the EEL spectra can be used to confirm the trapped water in the liquid cells[47,50–52]. The map in Figure 3a shows the localization of oxygen within the liquid cell. The extracted oxygen signal, integrated over the well, is shown in Figure 3b showing a signal to noise ratio (SNR~0.2) in line with previously reported trends for the dependence of SNR on liquid thickness.[51] We can also determine the presence of trapped liquid in the cells by measure their relative thickness using the log-ratio technique[49]. Figure 3c and 3d compares thickness mapping from filled and empty cells with the liquid filled cells having $t/\lambda \sim 0.3$ compared to $t/\lambda \sim 0.18$ for empty cells.

The high sensitivity of EELS to light elements makes it an ideal technique for the identification of water, however the presence of liquid limits SNRs for core loss mapping and we have found that EDXS provides more reliable elemental mapping of metal nanostructures in the EGLC liquid environment.

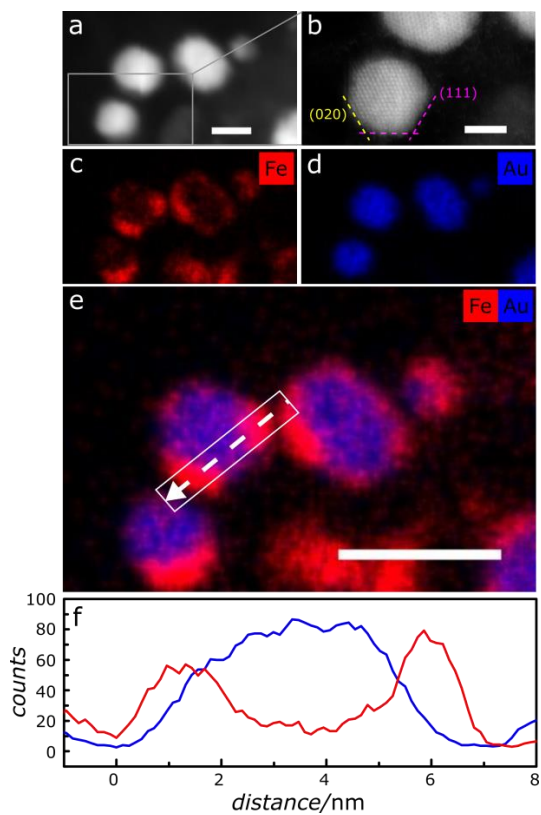

**Figure 4.** Elemental Imaging in a graphene liquid cell by STEM EDXS. (a) HAADF STEM image of FeAu nanoparticles simultaneously acquired with (c)-(e) Fe, Au and Fe+Au EDX elemental maps (80 kV). Note that the Fe shell was grown from solution in the liquid cell via beam-induced reduction on Au seed nanoparticles sputtered to the graphene window. EDX spectrum imaging was performed once the local iron concentration was depleted (complete reduction). The X-ray intensity profile (f) taken at the position shown on (e) reveals the spatial resolution of the elements, clearly resolving the ~1.5 nm thick Fe layer coating the Au core. (b) HAADF image of the same particles shown in (a) acquired later (imaged at 200kV to reveal atomic structure). Scale bars are (a) 5 nm, (b) 2.5 nm and (e) 10 nm

To demonstrate this we have performed highly spatially resolved EDXS mapping on Au nanoparticles in an aqueous solution containing Fe ions (for further details see SI). The widespread observation of oxygen was also observed in the EDX elemental maps (Figure S8) as expected for an aqueous environment.[48] The HAADF STEM image in Figure 4a shows typical Au nanocrystals with diameters 5.5-8.5 nm observed inside a liquid well (0.6 μm diameter cell, 30 nm deep). Aside from minimal motion within the first minute of imaging, no movement was observed for these particles, confirming they were attached to the graphene window (see SI). EDX spectrum imaging of the particles revealed a core-shell structure with Fe reduced from solution coating the surface of the Au seeds (Figure 4c-e). The complex particle geometry and low concentration of Fe means that this core-shell structure could not be determined unambiguously using the atomic number contrast present in HAADF STEM images as has been done for other bimetallic particles[41]. Analysis of the atomic resolution structure of the particle (Figure 4b) suggests that the Fe has been preferentially reduced on the Au nanoparticle vertices between (002) and (111) facets. The thickness of the iron surface layer varies from 1.5 – 2 nm, as illustrated by the line-scan in Figure 4f. The smallest interparticle gap can be measured as ~1 nm with the Fe X-ray signal reaching the background value in the middle of the scan. This elemental mapping capability is an order of magnitude better than previous state of the art liquid cell studies[48] and is equivalent to the best that is typically achievable for nanoparticle samples studied in vacuum.[53,54]

Puncturing the graphene window with the electron beam can be used to remove liquid from the cell as discussed previously (see Supplementary Video 2). We can then compare the effect on nanoparticle stability for prolonged spectrum imaging. When the liquid cells were 'emptied' of mobile liquid in this way, the bimetallic particles were no longer stable and quickly sintered during minutes of imaging, preventing acquisition of high quality elemental maps. EDX spectral imaging of the sintered structure shows migration of iron to the axial edges of the nanorod (further information see SI, Figure S10i). Similar sintering behavior has previously been observed to occur in gold and silver nanocrystals under ex situ electron beam illumination.[55] We attribute the unexpectedly high stability of the nanoparticles in the 'wet' liquid cells to the presence of the solution which may serve to passivate the nanocrystal surface and inhibit nanoparticle sintering.

In summary, we have designed and fabricated TEM-compatible engineered graphene liquid cells with controllable geometries, based on a lithographically patterned hBN spacer crystal with a specified thickness sandwiched between graphene windows. Unlike previous graphene liquid cells, our EGLC design is robust to vacuum cycling, and allows prolonged STEM imaging and analysis to be performed. The exceptional stability of the cells has enabled the first nanometer resolution elemental mapping of nanoparticles in liquid cells.

METHODS

AFM imaging was performed using a Bruker Dimension Icon AFM with a Nanoscope V controller, using Peakforce QNM imaging mode. Bruker ScanAsyst Air AFM tips with a nominal stiffness of 0.4 Nm$^{-1}$ were used, with a force setpoint of 2 nN to minimize tip induced membrane deformation. Images were flattened where necessary using Nanotec's WSXM software.

STEM Imaging and EDX spectroscopy analysis on the EGLCs was performed using a FEI Titan G2 80-200 S/TEM "ChemiSTEM" microscope operated at 80 kV to avoid knock-on damage of the graphene layers (unless noted otherwise). Imaging was carried out in HAADF STEM mode with a probe current of 20-180 pA for the Au/Fe nanoparticle study and 20-110 pA for the tungsten nanocrystal study and when imaging Pt nanocrystal formation, with a convergence semi-angle of 21 mrad in all cases. The dose rate in the tungsten nanocrystal study was $3.7 \times 10^5$ e$^-$/nm$^2$frame, calculated along the same lines as is presented in Abellan et al.[43] STEM images were recorded using FEI TIA software. Where possible, nanoparticles were imaged at cell edges to provide a reference for drift correction and minimize effects caused by the bowing of the graphene windows.

DualEELS was performed using a GIF Quantum ER System with an entrance aperture of 5 mm, 0.1s total dwell time and a dispersion of 0.25 eV/ch. EDX spectrum imaging was performed with a beam current of between 100 - 240 pA and acquisition times of between 2 and 20 minutes depending on the stability of the sample. All four of the Titan's Super-X SDD EDX detectors were used with a total collection solid angle of ~0.7 srad. EELS data was processed using Hyperspy[56] and EDX spectrum images processed using Bruker ESPRIT software.

## ASSOCIATED CONTENT

Description of fabrication procedures, description of vacuum cycling AFM study, supplementary STEM images, EELS spectra, EDX maps of nanoparticles, image processing and particle tracking methods (PDF)

Video 1 showing the dynamic motion and growth of nanoparticles in a liquid environment. Video runs at 20 times real-time (AVI)

Video 2 showing a leaking cell with liquid flow. Video runs at 20 times real-time (AVI)

## AUTHOR INFORMATION

**Corresponding Author**

*E-mail: sarah.haigh@manchester.ac.uk
*E-mail: roman@manchester.ac.uk

**Author Contributions**

D.J.K, N.C, R.V.G and S.J.H. wrote the manuscript. M.Z, N.C, M.J.H. and R.V.G. fabricated the liquid cells and performed AFM measurements. D.J.K., E.A.L. and S.J.H. performed STEM characterization. A.M.R. performed FIB-SEM characterization. All authors contributed to the interpretation of the results and have given approval to the final version of the manuscript.

**Notes**

The authors declare no competing financial interest.

ACKNOWLEDGMENT

The authors acknowledge funding support from the Engineering and Physical Sciences Research Council (EPSRC) UK Grants # EP/G035954/1, EP/K016946/1, EP/M010619/1 and EP/J021172/1, the EPSRC Graphene NowNano Doctoral Training center, Defense Threat Reduction Agency grant HDTRA1-12-1-0013, ERC Starter Grant (EvoluTEM) and the Royal Society University Research Fellowship Scheme.

Eljarrat, A.; Mazzucco, S.; Migunov, V.; Aarholt, T.; Walls, M.; Winkler, F.; Martineau, B.; Donval, G.; Hoglund, E. R.; Alxneit, I.; Hjorth, I.; Zagonel, L. F.; Garmannslund, A.; Gohlke, C.; Iyengar, I.; Chang, H.-W. **2017**.